\documentclass[journal,a4paper,onecolumn]{IEEEtran}

\usepackage{epsfig}
\usepackage{epstopdf}
\usepackage{siunitx}
\usepackage{pbox}
\usepackage{makecell}
\usepackage{multirow}
\usepackage{amssymb,amsmath}
\usepackage[normalem]{ulem}
\usepackage{gensymb}
\ifCLASSOPTIONcompsoc
\usepackage[caption=false,font=normalsize,labelfon
t=sf,textfont=sf]{subfig}
\else
\usepackage[caption=false,font=footnotesize]{subfi
	g}
\fi 
\usepackage{cite}

\begin{document}
%\IEEEpubid{0000--0000/00\$00.00˜\copyright˜2007 IEEE}

\title{Multifunctional high-frequency circuit capabilities of ambipolar carbon nanotube FETs}

\author{Javier N. Ramos-Silva, An\'ibal Pacheco-S\'anchez, Eloy Ram\'irez-Garc\'ia and David Jim\'enez 
\thanks{This project has received funding from Instituto Polit\'ecnico Nacional under the contract no. SIP/20210167, from the European Union's Horizon 2020 research and innovation programme under grant agreements No GrapheneCore2 785219 and No GrapheneCore3 881603, from Ministerio de Ciencia, Innovaci\'on y Universidades under grant agreement RTI2018-097876-B-C21(MCIU/AEI/FEDER, UE). This  article  has been partially  funded  by  the  European  Regional  Development  Funds  (ERDF)  allocated  to  the  Programa Operatiu FEDER de Catalunya 2014-2020, with the support of the Secretaria d'Universitats i Recerca of the Departament d'Empresa i Coneixement of the Generalitat de Catalunya for emerging technology clusters to  carry  out  valorization  and  transfer  of  research  results.  Reference  of  the  GraphCAT  project:  001-P-001702.}
\thanks{Javier N. Ramos-Silva is with Instituto Polit\'ecnico Nacional, UPALM, Edif. Z-4 3er Piso, Cd. de M\'exico, 07738, M\'exico and also with the Departament d'Enginyeria Electr\`onica, Escola d'Enginyeria, Universitat Aut\`onoma de Barcelona, Bellaterra 08193, Spain, e-mail: jramoss1303@alumno.ipn.mx}
\thanks{An\'ibal Pacheco-S\'anchez and D. Jim\'enez are with the Departament d'Enginyeria Electr\`onica, Escola d'Enginyeria, Universitat Aut\`onoma de Barcelona, Bellaterra 08193, Spain, e-mails: AnibalUriel.Pacheco@uab.cat, David.Jimenez@uab.cat}
\thanks{E. Ram\'irez-Garc\'ia is with Instituto Polit\'ecnico Nacional, UPALM, Edif. Z-4 3er Piso, Cd. de M\'exico, 07738, M\'exico, e-mail: ramirezg@ipn.mx}
% <-this % stops a space
}
\maketitle
\makeatletter
%\makeatletter
\def\ps@IEEEtitlepagestyle{
  \def\@oddfoot{\mycopyrightnotice}
  \def\@evenfoot{}
}
\def\mycopyrightnotice{
  {\footnotesize
  \begin{minipage}{\textwidth}
  \centering
© 2021 IEEE.  Personal use of this material is permitted.  Permission from IEEE must be obtained for all other uses, in any current or future media, including reprinting/republishing this material for advertising or promotional purposes, creating new collective works, for resale or redistribution to servers or lists, or reuse of any copyrighted component of this work in other works.
  \end{minipage}
  }
}

\markboth{}
\MakeLowercase

\begin{abstract}
\boldmath
An experimentally-calibrated carbon nanotube compact transistor model has been used here to design two high-frequency (HF) circuits with two different functionalities each: a phase configurable amplifier (PCA) and a frequency configurable amplifier (FCA). The former design involves an in-phase amplifier and an inverting amplifier while the latter design embraces a frequency doubler as well as a distinct inverting amplifier. The specific functionality selection of each of the two HF circuit designs is enabled mainly by the inherent ambipolar feature at a device level. Furthermore, at a circuit level the matching networks are the same regardless the operation mode. In-phase and inverting amplification are enabled in the PCA by switching the gate-to-source voltage ($V_{\rm GS}$) from \SI{-0.3}{\volt} to \SI{0.9}{\volt} while the drain-to-source voltage ($V_{\rm DS}$) remains at \SI{3}{\volt}. By designing carefully the matching and stability networks, power gains of $\sim$\SI{4.5}{\decibel} and $\sim$\SI{6.7}{\decibel} at \SI{2.4}{\giga\hertz} for the in-phase and inverting operation mode have been achieved, respectively. The FCA, in its frequency doubler operation mode, exhibits $\sim$\SI{20}{\decibel c} of fundamental-harmonic suppression at \SI{2.4}{\giga\hertz} when an input signal at \SI{1.2}{\giga\hertz} is considered. This frequency doubler functionality is enabled at $V_{\rm GS}=\SI{0.3}{\volt}$, whereas at $V_{\rm GS}=\SI{0.9}{\volt}$ amplification of $\sim$\SI{4.5}{\decibel} is obtained while $V_{\rm DS}$ remains at \SI{3}{\volt} in both cases. In both configurable circuits the stabilization and matching networks are the same regardless the bias-chosen operation mode. The circuits performance degradation due to metallic tubes in the device channel is studied as well as the impact of non-ideal inductors in each design. PCA and FCA operation modes are further exploited in high-frequency modulators.
\end{abstract}


\IEEEpeerreviewmaketitle

%\pacs{}% insert suggested PACS numbers in braces on next line

%\maketitle %\maketitle must follow title, authors, abstract and \pacs

% References should be done using the  \cite, \ref, and \label commands
\section{Introduction}
\label{ch:intro}

Carbon nanotubes (CNTs) have emerged as one of the options to replace the silicon (Si)-based channel of novel field-effect transistors (FETs) in high-performance applications beyond the limits of conventional technologies \cite{Mar14}, \cite{HarHer21} while improving further the device footprint and fabrication costs of related circuits \cite{QiuZha17}. The quasi-ballistic transport, an outstanding gate control over the channel, as well as an inherent device linearity associated to CNTFETs, makes them suitable for low-power high-frequency (HF) applications \cite{HarHer21}, \cite{MotCla15}, \cite{Maa17} as recently demonstrated by devices operating at frequencies of around \SI{100}{\giga\hertz} \cite{CaoBra16,RutKan19,ZhoShi19}. One feature of carbon-based FETs is their ambipolar behavior \cite{MarDer01,RadHei03,JavGuo04,ZhaWan09}, i.e., the device performance in the active region can be of \textit{n}-type-like or \textit{p}-type-like according to the bias. Notice that CNTFETs in this work have only three terminals (source, drain and gate), in contrast to previous reports where an additional control gate has been used to control the ambipolarity \cite{BenJMoh11}, \cite{MouTie18}. 

In-phase and inverting amplification can be enabled in ambipolar transistors by applying an input AC signal and biasing the device in the linear \textit{n}-type region and linear \textit{p}-type region, respectively. Furthermore, if the DC device operation is around the minimum transconductance point, the frequency ($f$) of an input AC signal can be multiplied at the output \cite{WanZha13}. Therefore, on-chip designs can benefit from ambipolar devices since stages of radio-frequency (RF) integrated circuits, including frequency multipliers and amplifiers, can be fabricated with the same transistor technology biased at different regions, hence reducing the fabrication process and production costs.

CNTFETs ambipolarity \cite{RadHei03}, \cite{PacFuc18,AppRad04,LinApp04,JimCar06} has been already used in proof-of-concept analog circuits \cite{WanZha13,WanLia14,WanDin10,CheLin13} focused on the frequency multiplying characteristic rather than tackling efficient circuit design to exploit the device multifunctionality for analog circuits. In this work, two CNTFET-based HF circuit designs, identified here as a phase configurable amplifier (PCA) and a frequency configurable amplifier (FCA), are presented. The device biasing conditions enable two different circuit operation modes in each design: an in-phase amplifier (PA) and an inverting amplifier (IA) are possible in the PCA while a frequency doubler (FD) and a distinct IA can be selected in the FCA. The different functionalities for the PCA and the FCA have been demonstrated with identical matching networks and further exploited for the proposal of CNTFET-based modulators. Hence, this is an efficient on-chip solution for CNTFET-based configurable HF analog circuits by using a suitable compact large-signal model \cite{SchHaf15}.

This work intends to show the capabilities of ambipolar CNTFETs in high-frequency circuits by considering optimal and quasi-optimal conditions at both device and circuit level. The outcome of this study should be considered as a support and motivation for the CNTFET community related to fabrication and circuit design since these results have been obtained considering feasible scenarios.

\section{Compact model and device description}
\label{ch:mod}

\begin{figure}[!hbt]
	\centering
	{\includegraphics[height=0.22\textwidth]{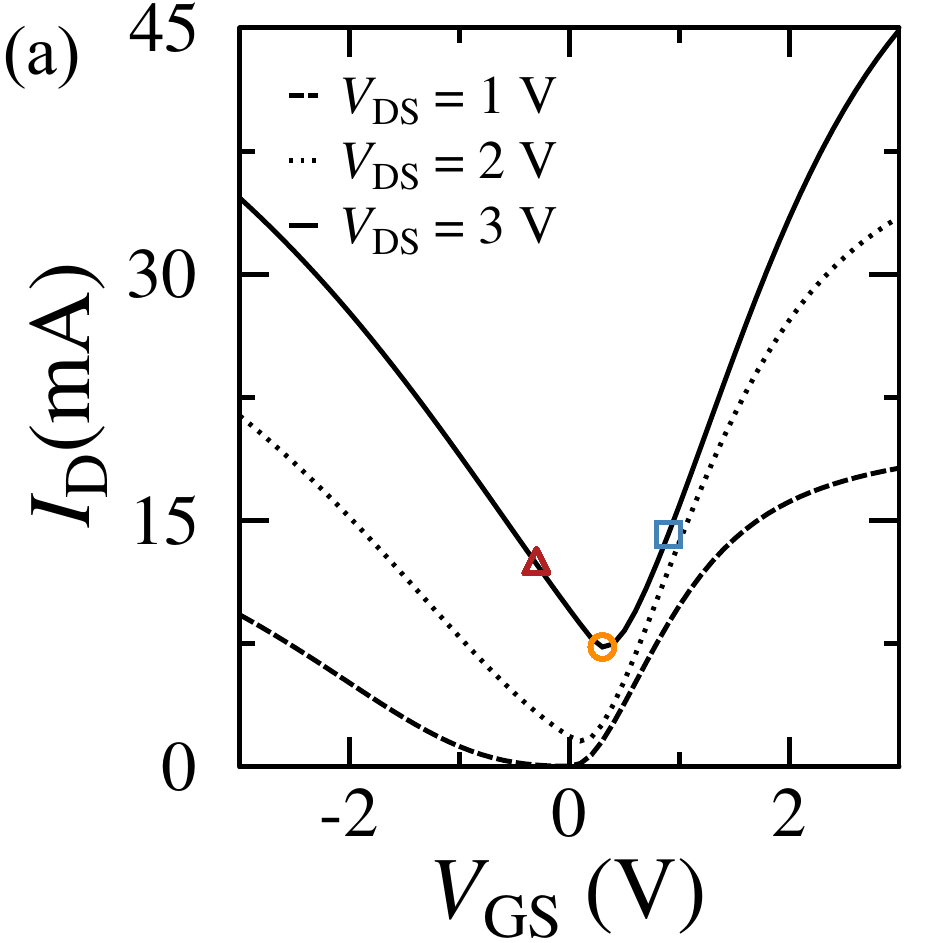}}
	{\includegraphics[height=0.22\textwidth]{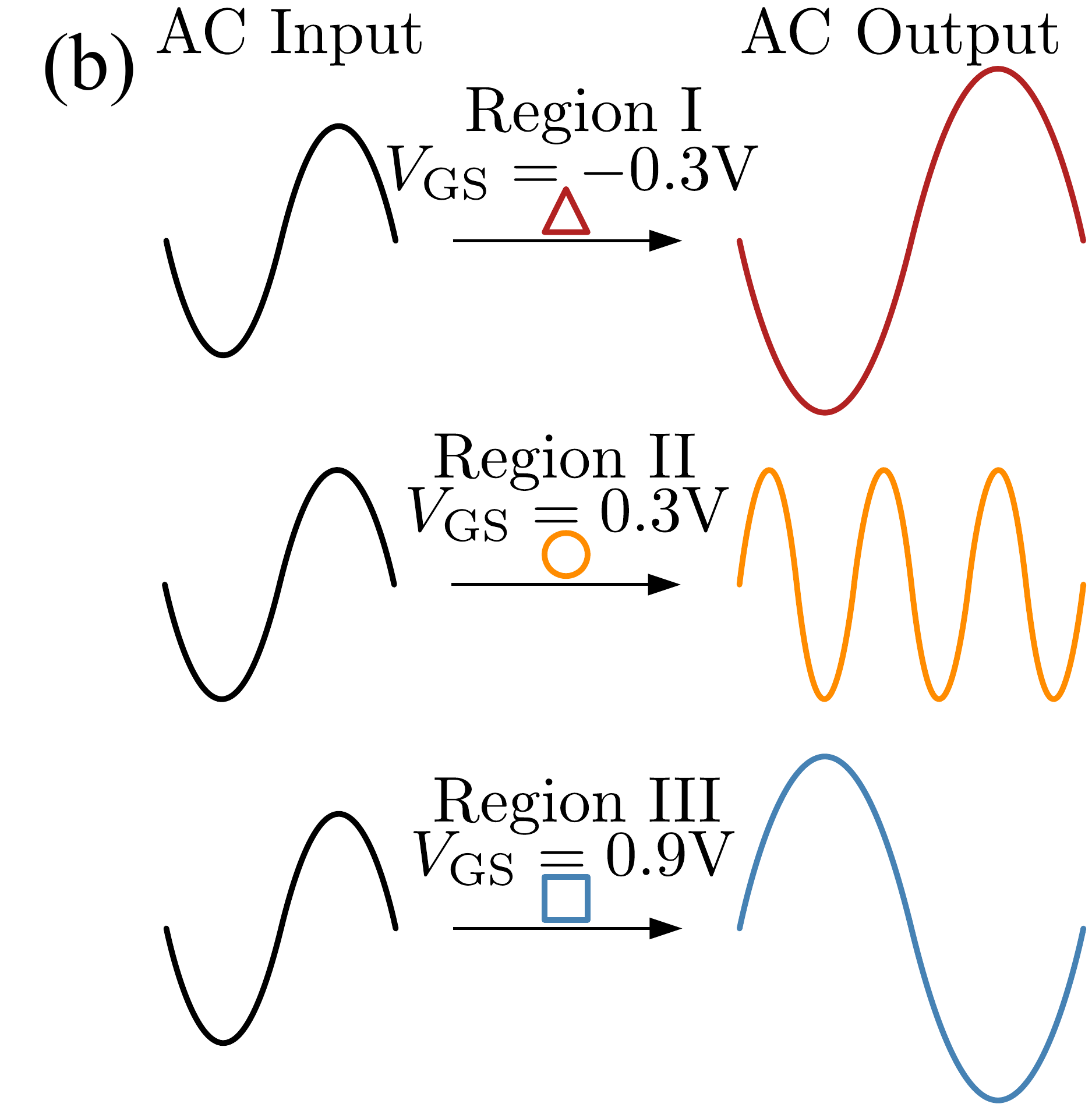}}
	\caption{(a) Simulated transfer characteristics from CCAM for the reference MT-CNTFET without metallic tubes. Markers indicate the different bias points considered in this work. (b) Representation of the  $v_{\rm in}$ (black) and ideal $v_{\rm out}$ (colored) for the device biased at the different operation regions.}
	\label{fig:Transfer}
\end{figure}

Electron and hole transport is inherently allowed simultaneously in CNTs. The dominant carrier type in CNTFETs is determined by the contacts physics \cite{RadHei03}, \cite{AppRad04}, \cite{PacFuc18}. In the transfer curves shown in Fig. \ref{fig:Transfer}(a), the drain current ($I_{\rm D}$) is comprised mainly by tunneling electron current at the saturation regime of the \textit{n}-branch, due to a strong energy bands bending (thin source potential barrier). The ohmic region of the \textit{p}-branch is due to a thin drain barrier induced by high negative gate-to-source voltage $V_{\rm GS}$ enabling a major contribution of tunneling hole current. 

\begin{figure}[!hbt]
	\centering
	{\includegraphics[height=0.28\textwidth]{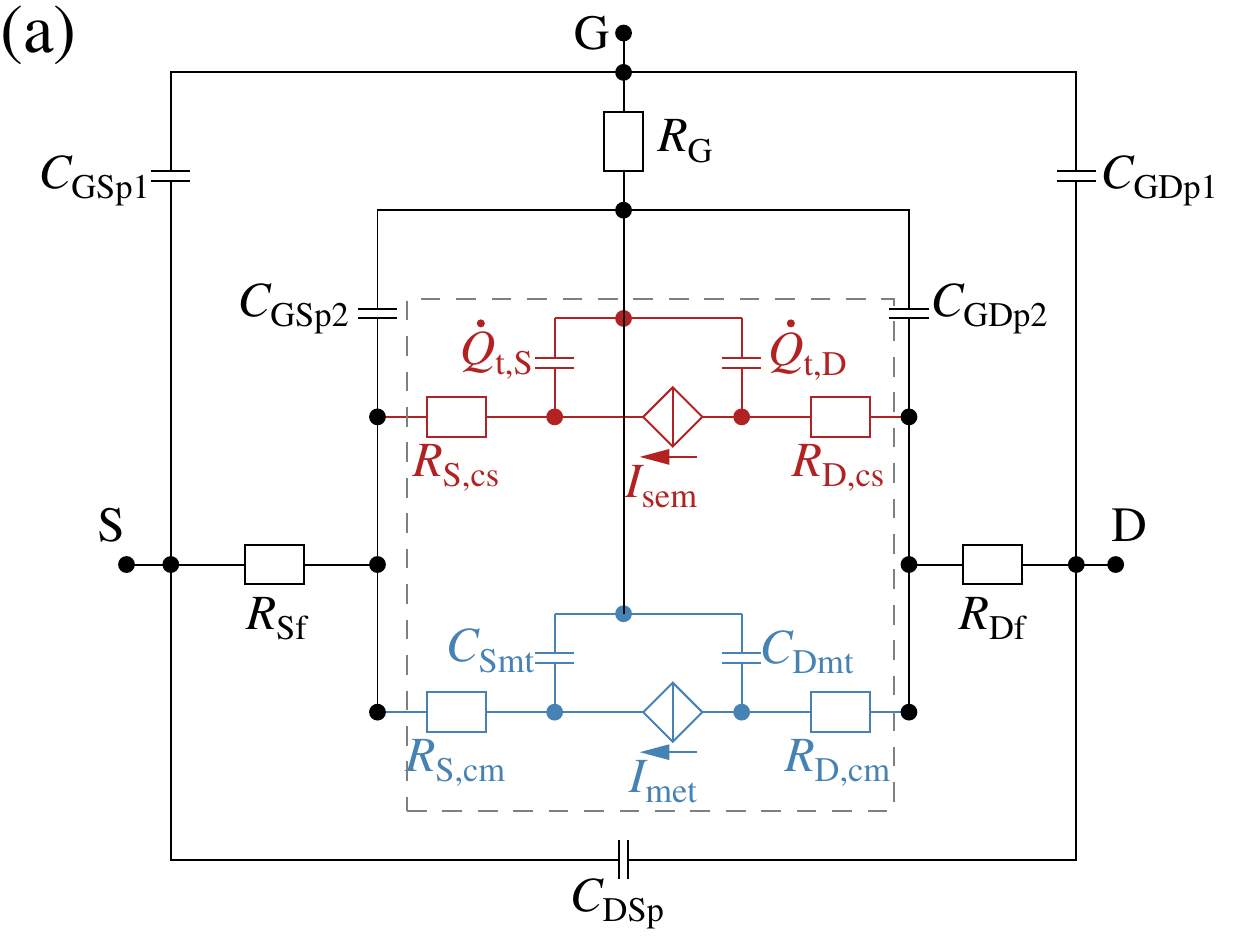}}
	{\includegraphics[height=0.16\textwidth]{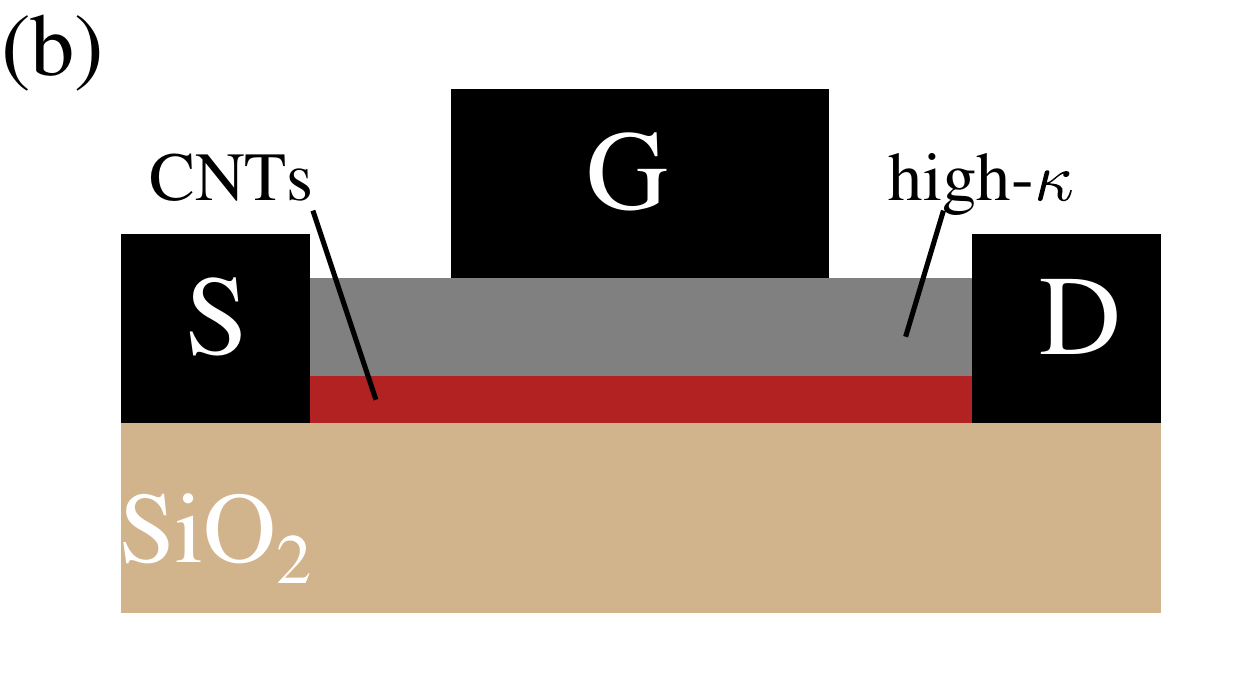}}
	\caption{(a) Equivalent circuit of the large-signal model CCAM. The intrinsic part (inside the dashed box) includes a module for parameters associated to semiconducting tubes (in red in the electronic version) and another one related to metallic tubes (in blue in the electronic version). See \cite{SchHaf15,SchHaf15_2} for detailed discussions on each parameter as well as for the experimental validation of the model. (b) Schematic cross-section of the CNTFET used for the CCAM calibration used in this work (not drawn to scale).}
	\label{fig:ccam}
\end{figure}

In this work, a semi-physical large-signal compact model for CNTFETs, named CCAM \cite{SchHaf15,SchHaf15_2} has been used. It describes the ambipolar behavior of a fabricated CNTFET technology considering, by different modules, both the contribution of semiconductor (s-) and metallic (m-) tubes in the channel. These modules correspond to the intrinsic part of the device as identified in the compact model equivalent circuit shown in Fig. \ref{fig:ccam}(a). Discussions on each model parameter and the physics behind each of them have been provided in \cite{SchHaf15,SchHaf15_2}. High-frequency noise modules, described elsewhere \cite{RamPac19} have been activated in this work. The compact model has been calibrated (including both intrinsic part and parasitics) to hysteresis-free DC and dynamic experimental data \cite{SchHaf15}, \cite{HafCla14} from a fabricated top-gate multi-tube (MT) CNTFET technology \cite{SchKol11}, hence, the model used here is directly related to a manufacturable optimized device technology\footnote{The compact model used here is one of the few able to correctly describe the performance of fabricated RF CNTFET technology}. CNTFET-based RF applications have been already developed using this technology \cite{SchKol11,EroLin11,TagCar15,TagCar17}. HF performance projections for amplifiers using the considered CNTFET technology have been developed using CCAM \cite{ClaMuk14,RamPac19}, however, the ambipolarity features of CNTFETs have not been exploited in such works. The reference device considered here has a channel length of \SI{700}{\nano\meter} and a top-gate length of \SI{250}{\nano\meter}, a total of eight gate fingers, \SI{50}{\micro\meter} of width each, yielding a device total width of \SI{400}{\micro\meter}. The device cross-section is shown in Fig. \ref{fig:ccam}(b). Diameters of the tubes have been estimated of $\sim$\SI{1.8}{\nano\meter} \cite{SchHaf15}, leading to a channel bandgap of $\sim$\SI{0.47}{\electronvolt}. The channel array has s- and m-tubes grown on SiO$_{2}$ substrate via chemical vapor deposition. Fabrication process details can be found in \cite{SchKol11}.
%The fabrication process of this technology can be found in detail in \cite{SchKol11}. 

A high number of s-CNTs in the channel is required towards improving the HF device performance, as demonstrated by an s-tube sensitivy study for RF circuits presented elsewhere \cite{ClaMuk14} and fabricated RF state-of-the-art devices \cite{CaoBra16,RutKan19,CaoChe16,ZhoShi19}. Hence, the number of m-CNTs in the compact model has been set to zero while other parameters remain unchanged for enhancing the HF performance, i.e., an optimized device with only \SI{2000}{} s-CNTs in the channel, leading to a tube density $n_{\rm t}$ of \SI{5}{} CNTs$/\SI{}{\micro\meter}$, has been considered in this study. Transfer characteristics of the simulated ambipolar MT-CNTFET, considering only s-tubes in the channel, are shown in Fig. \ref{fig:Transfer}(a) where three bias points, related to a different operation region each, have been indicated. Region I corresponds to the \textit{p}-type behavior, region II embraces the polarity transition (low-transconductance regime) and region III includes the \textit{n}-type dominant branch. At high $V_{\rm DS}$, tunneling processes are more pronounced as observed by the higher value of the minimum current at the corresponding transfer characteristic in contrast to curves at lower $V_{\rm DS}$. The expected outcome of an input AC signal for the device biased differently is qualitatively shown in Fig.\ref{fig:Transfer}(b). 

\section{CNTFET-based multifunctional circuits}
\label{ch:mul}

The ambipolar behavior of CNTFETs enhances the development of multifunctional analog circuits by biasing the device in one of the three regions identified in Fig. \ref{fig:Transfer}. In regions I and III, the dependence of $I_{\rm D}$ on $V_{\rm GS}$ is ideally described by a linear function with a negative (region I) or positive (region III) slope, i.e., ${\textstyle I_{\rm D}=-g_{\rm m}V_{\rm GS}+B}$ or ${\textstyle I_{\rm D}=g_{\rm m}V_{\rm GS}+E}$, respectively. In region II, ${\textstyle I_{\rm D}}$ has a parabolic-like $V_{\rm GS}$-dependence, i.e., ${\textstyle I_{\rm D}=C+D(V_{\rm GS}-V_{\rm GS_0})^2}$, with the constants $B$, $C$, $D$ and $E$ and ${\textstyle V_{\rm GS_0}=V_{\rm GS}\vert_{\rm{min}(\mathit{I}_{\rm D}\rm )}}$. Thus, considering ${\textstyle V_{\rm in}=V_{\rm GS}+v_{\rm in}=V_{\rm GS}+A\sin \omega t }$, in regions I and III, the device operates as a $g_{\rm m}$-controlled amplifier: being an in-phase amplifier for ${\textstyle g_{\rm m}<0}$ (in region I), while in region III (${\textstyle g_{\rm m}>0}$) the device works as an inverting amplifier. An improved device linearity (constant $g_{\rm m}$ \cite{MotCla15}) enhance the amplifying performance. Furthermore, if the device is biased in region II, it works during half period of the input signal in the $p$-like region and during the other half period in the $n$-like region, yielding a doubled frequency in the output signal enhanced by the parabolic-shape transfer characteristics within this bias. Due to the common source topology, by using a drain resistor $R_{\rm D}$ to apply $V_{\rm DS}$ from a $V_{\rm DD}$ voltage source (see Fig. \ref{fig:Pashe_Quadra}), the output voltage in region II is ${\textstyle V_{\rm out}=V_{\rm DS}+v_{\rm out}=}$ ${\textstyle V_{\rm DD}-BR_{\rm D}-1/2\left(R_{\rm D}CA^2\right)+1/2\left(R_{\rm D}CA^2\cos 2\omega t \right)}$, i.e., the doubled frequency is explained by the fourth \-term. A similar analysis of the output signals for different operation modes of ambipolar CNTFETs has been provided elsewhere \cite{WanZha13}. 

%\vspace{-0.1cm}
\subsection{Phase configurable amplifier design}

The selected bias points of the PCA are ${\textstyle V_{\rm GS}=-\SI{0.3}{\volt}}$ (region I) for the PA operation mode and ${\textstyle V_{\rm GS}=\SI{0.9}{\volt}}$ (region III) for the IA operation mode, in both cases ${\textstyle V_{\rm DS}=\SI{3}{\volt}}$. Cutoff and maximum oscillation frequencies obtained at the bias point in region I(III) are \SI{4}{\giga\hertz}(\SI{8.5}{\giga\hertz}) and \SI{14.6}{\giga\hertz}(\SI{23.5}{\giga\hertz}), respectively. Both amplifiers have been designed for maximum gain at \SI{2.4}{\giga\hertz}. The device unilateral power gain at \SI{2.4}{\giga\hertz} are \SI{19}{\decibel} and \SI{10}{\decibel} for the PCA-PA and PCA-IA, respectively. Ambipolar CNTFETs have been used previously to design in-phase and inverting amplifiers at low operation frequencies \cite{WanZha13}, biasing the device at high $V_{\rm DD}$ (\SI{12}{\volt}), lacking gain for the in-phase configuration and requiring specific matching networks for each operation mode. In contrast, in this work the considered device is used to design a PCA suitable for RF applications at lower $V_{\rm DD}$ ($=V_{\rm DS}=\SI{3}{\volt})$ and with identical matching and stabilization networks regardless the operation mode, i.e., a more efficient design for multifunctional CNTFET-based circuits is proposed here.  

\begin{figure}[!hbt]
	\centering
	\includegraphics[height=0.22\textwidth]{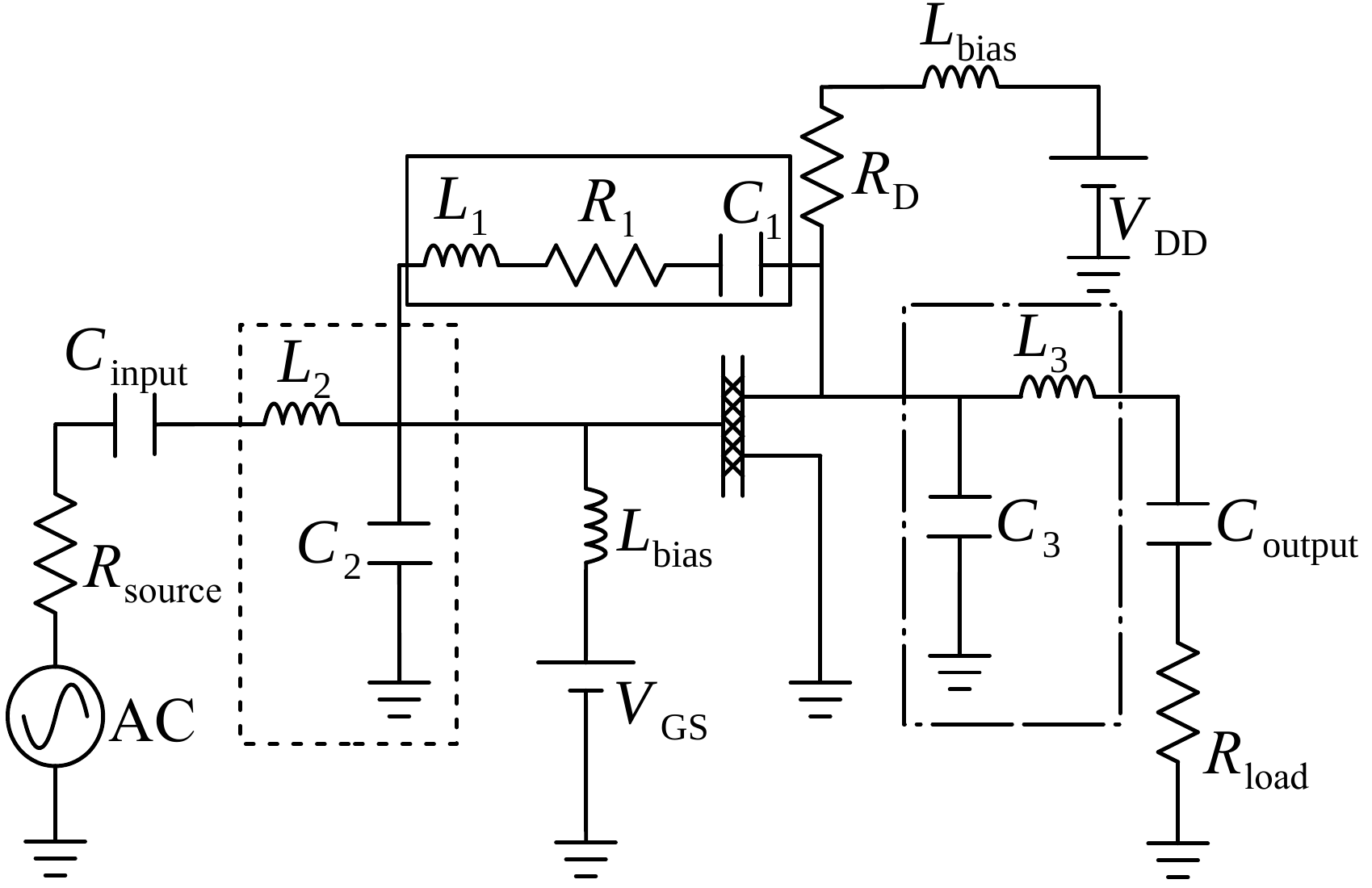}
	\caption{Schematic of the PCA and FCA. Possible operation modes for the PCA (${\textstyle R_{\rm D}=\SI{0}{\ohm}}$) are as an in-phase or as an inverting amplifier. The FCA (${\textstyle R_{\rm D}\neq\SI{0}{\ohm}}$) can operate as a frequency doubler or as an inverting amplifier. All operation modes are controlled by $V_{\rm GS}$ in each configuration. Matching (dotted and dash-dotted line boxes) and stabilization (continuous line box) networks are the same regardless the operation mode of each design. Parameter values are reported in Table \ref{Values}.}
	\label{fig:Pashe_Quadra}
\end{figure}

\begin{figure}[!hbt]
	\centering
	{\includegraphics[height=0.22\textwidth]{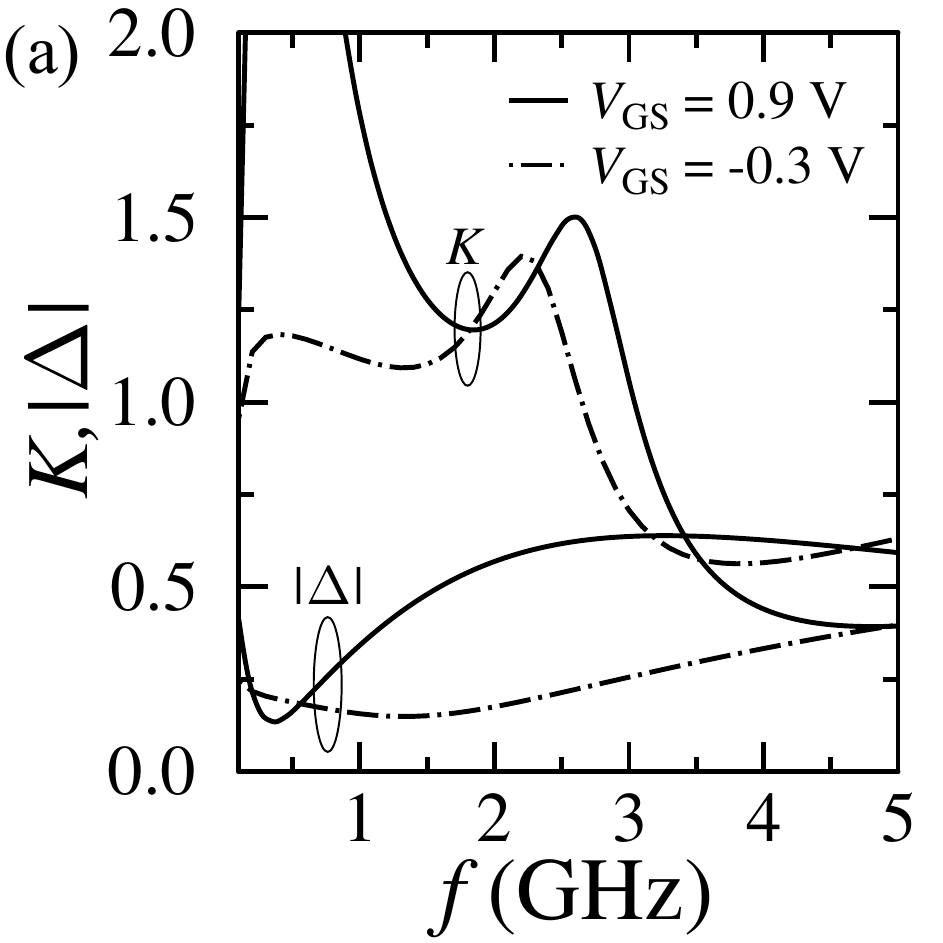}}
	{\includegraphics[height=0.22\textwidth]{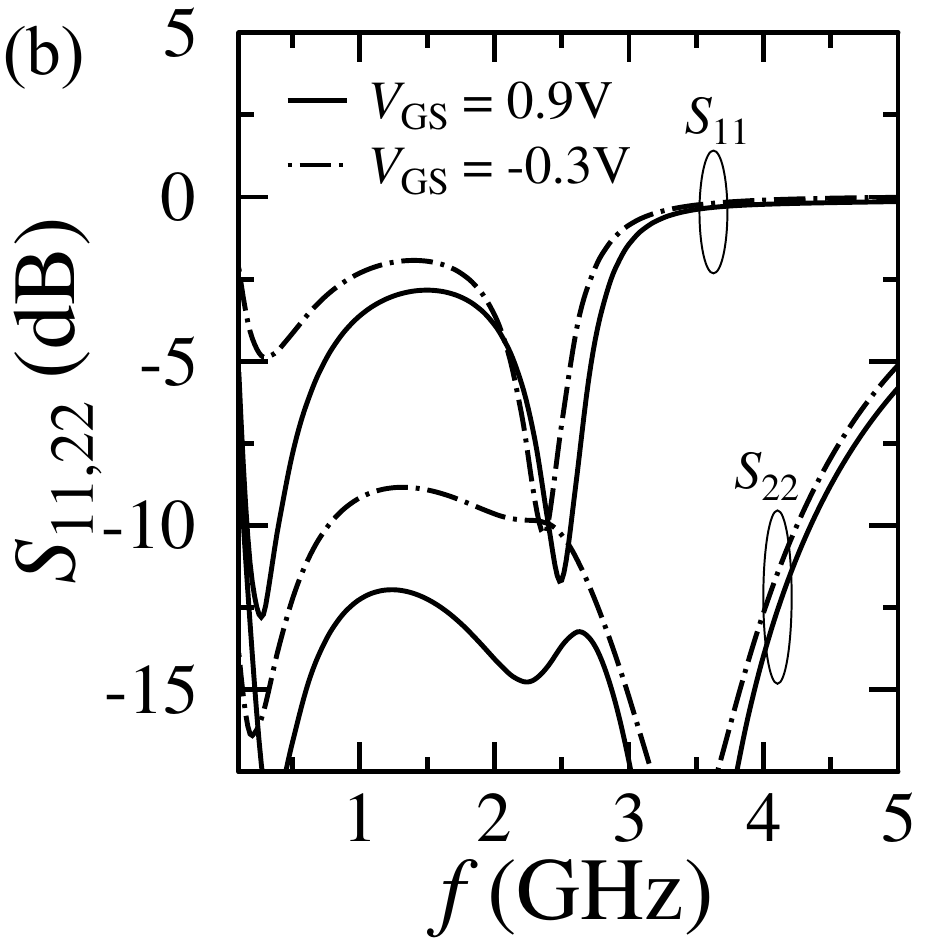}}
	\caption{(a) Stability metrics and (b) $\vert S_{11}\vert$ and $\vert S_{22}\vert$ for PCA operation modes: in-phase amplifier (${\textstyle V_{\rm GS}=\SI{-0.3}{\volt}}$) and inverting amplifier (${\textstyle V_{\rm GS}=\SI{0.9}{\volt}}$).}
	\label{fig:K}
\end{figure}

Fig. \ref{fig:Pashe_Quadra} shows the proposed PCA design operating at a $f$ of \SI{2.4}{\giga\hertz}. The parameters values are listed in Table \ref{Values}(a). The design has been developed using the Verilog-A implementation \cite{SchHaf15_2} of CCAM \cite{SchHaf15} into the Keysight Advance Design System software. Based on the internal transport mechanisms at each contact, source and drain contacts remain physically the same regardless the operation mode. However, at a circuit level and considering the difference in the external potentials, a common-source (common-drain) circuit topology is related for the CNTFET working in region I (III). A feedback topology using a resistor $R_1$, a capacitor $C_1$ and an inductor $L_1$ has been designed in order to find stability conditions (${\textstyle K>1}$ and ${\textstyle \vert\Delta\vert<1}$, see \cite{Poz11}). The PCA is unconditionally stable at \SI{2.4}{\giga\hertz}, and up to \SI{3}{\giga\hertz}, while $V_{\rm GS}$ changes indistinctly the operation mode (see Fig. \ref{fig:K}(a)). Furthermore, matching networks have been designed towards a transparent implementation with established \SI{50}{\ohm}-RF systems. Notice that the stability and matching networks remain the same regardless the operation mode (bias). For higher frequencies however, matching and stability network designs, different to the ones considered here, are required to avoid non-desired oscillations. 

%\vspace{-0.5cm}
\begin{table} [!hbt] 
	\centering 
	\caption{Parameter values of the designed circuits}
	\subfloat[][PCA]{
		\centering
		\begin{tabular}{c||c}
			Parameter & Value \\ \hline \hline
            $V_{\rm DD}$&\makecell{\SI{3}{\volt} (PA)\\ \SI{3}{\volt} (IA)}  \\  \hline			
			$V_{\rm DS}$& \SI{3}{\volt} \\ \hline
			$V_{\rm GS}$&\makecell{-\SI{0.3}{\volt} (PA)\\ \SI{0.9}{\volt} (IA)}  \\  \hline
			$R_{\rm D}$&\SI{0}{\ohm} \\ \hline
			$R_{1}$&\SI{100}{\ohm}\\ \hline
			$C_1$&\SI{10}{\pico\farad}\\ \hline
			$L_1$&\SI{20}{\nano\henry}\\ \hline
			$L_2$&\SI{12.6}{\nano\henry}\\ \hline
			$C_2$&\SI{124.5}{\femto\farad}\\ \hline
			$L_3$&\SI{2.6}{\nano\henry}\\ \hline
			$C_3$&\SI{419.1}{\femto\farad}\\	
			\hline
		\end{tabular} 
	}
	\hspace{0.0cm}
	\subfloat[][FCA]{
		\centering
		\begin{tabular}{c||c}
			Parameter & Value \\ \hline \hline
            $V_{\rm DD}$&\makecell{\SI{4}{\volt} (FD)\\ \SI{5}{\volt} (IA)}  \\  \hline			
			$V_{\rm DS}$& \SI{3}{\volt} \\ \hline
		$V_{\rm GS}$&\makecell{\SI{0.3}{\volt} (FD)\\ \SI{0.9}{\volt} (IA)}  \\  \hline
			$R_{\rm D}$&\SI{137}{\ohm} \\ \hline
			$R_{1}$&\SI{330}{\ohm}\\ \hline
			$C_1$&\SI{10}{\pico\farad}\\ \hline
			$L_1$&\SI{45}{\nano\henry}\\ \hline
			$L_2$&\SI{24.8}{\nano\henry}\\ \hline
			$C_2$&\SI{472.7}{\femto\farad}\\ \hline
			$L_3$&\SI{4}{\nano\henry}\\ \hline
			$C_3$&\SI{522}{\femto\farad}\\
			\hline
		\end{tabular}
	}
	\label{Values}
\end{table}
%\vspace{-0.5cm}

\begin{figure}[!hbt]
	\centering
	\includegraphics[height=0.22\textwidth]{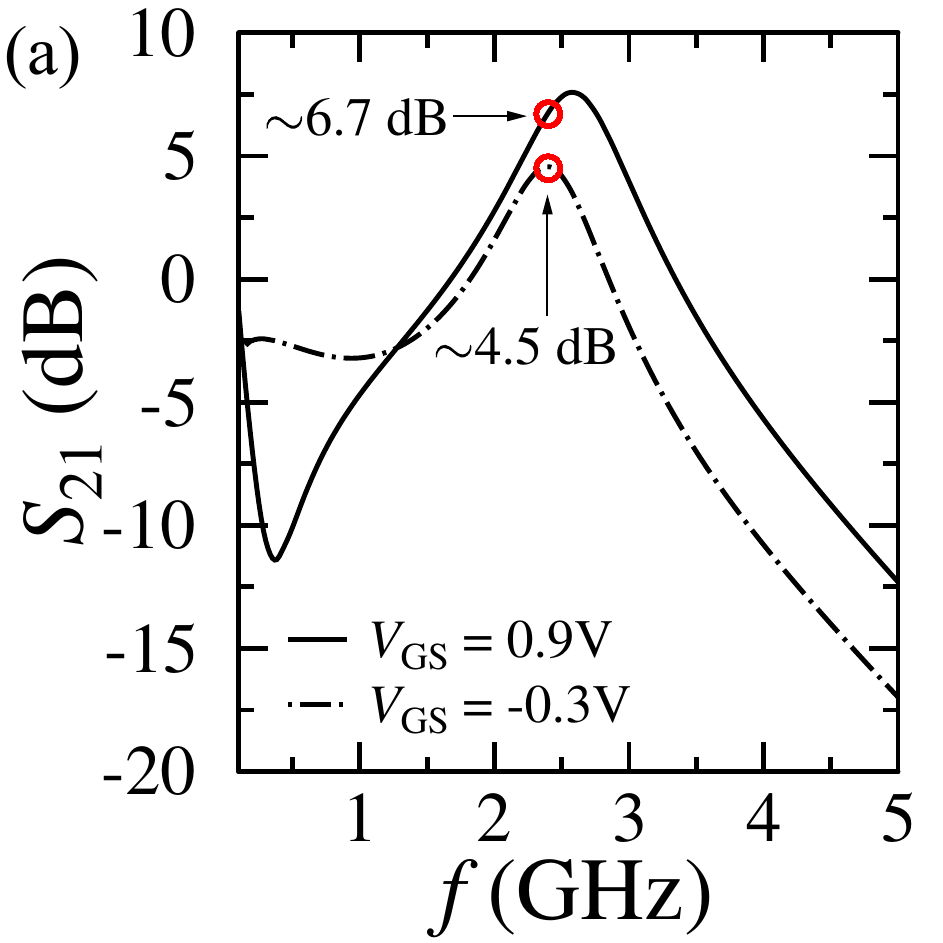}
	\includegraphics[height=0.22\textwidth]{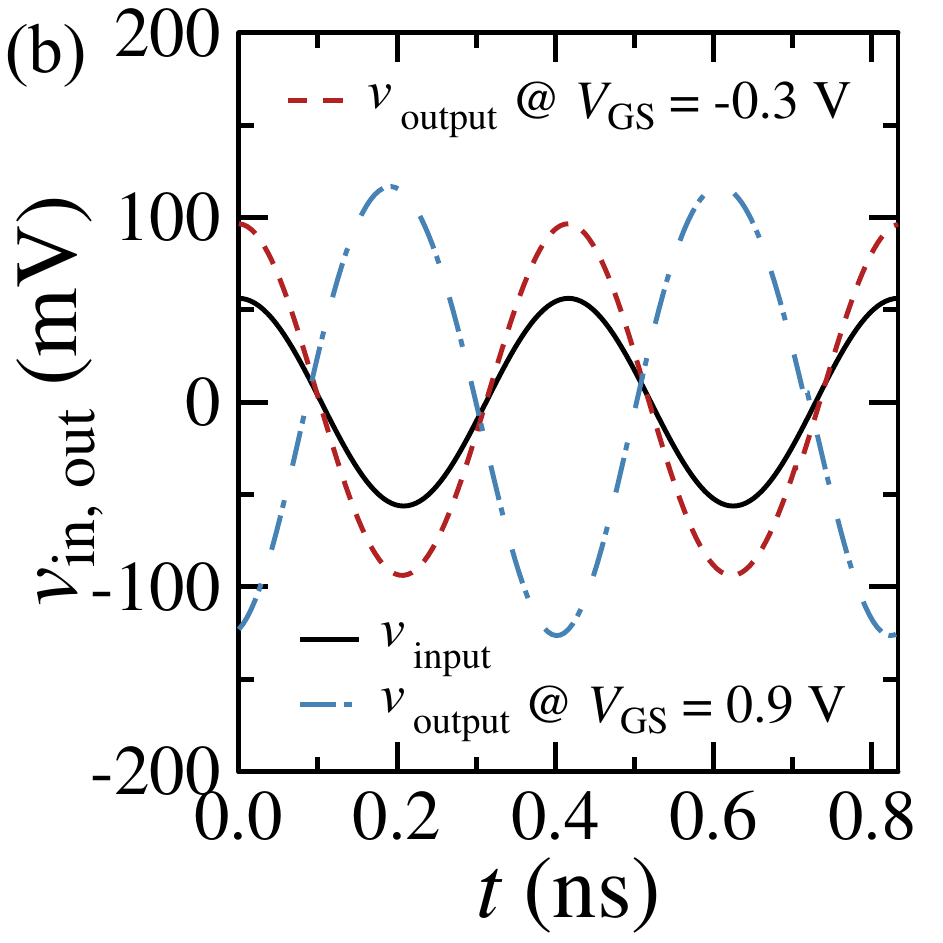}
	\caption{(a) $\vert S_{21}\vert$ and (b) $v_{\rm in}$ and $v_{\rm out}$ signals for the two PCA operation modes: in-phase amplifier ($V_{\rm GS}=\SI{-0.3}{\volt}$) and inverting amplifier ($V_{\rm GS}=\SI{0.9}{\volt}$). Circles in (a) indicate the gain at \SI{2.4}{\giga\hertz}. (a slight phase shift caused by the matching networks, the same in both operation modes, has been corrected for a better visualization of the circuit performance).}
	\label{fig:S21_Pashe_Quadra}
\end{figure}

The PCA is able to change its operation mode only by switching the device bias between regions I and III. In both cases, $\vert S_{11}\vert$ and ${\textstyle \vert S_{22}\vert\lesssim -\SI{10}{\decibel}}$ at \SI{2.4}{\giga\hertz} as shows Fig. \ref{fig:K}(b). The power gain for the in-phase amplifier operation mode is $\sim$\SI{4.5}{\decibel} whereas for the inverting amplifier mode is $\sim$\SI{6.7}{\decibel} (see Fig. \ref{fig:S21_Pashe_Quadra}(a)). In contrast to voltage-gainless fabricated CNTFET-based ambipolar amplifiers designed for specific functionalities \cite{WanZha13}, voltage gain of $\sim$\SI{1.7}{\volt/\volt} and $\sim$\SI{2.16}{\volt/\volt} has been obtained here, for the in-phase and inverting amplifier, respectively. By switching operation modes, the amplified output signal phase is shifted from $0\degree$ (${\textstyle V_{\rm GS}=\SI{-0.3}{\volt}}$) to $180\degree$ (${\textstyle V_{\rm GS}=\SI{0.9}{\volt}}$) (see Fig. \ref{fig:S21_Pashe_Quadra}(b)). The input power ($P_{\rm in}$) has been of $-\SI{15}{\decibel m}$ at \SI{2.4}{\giga\hertz}.

In order to characterize entirely the proposed amplifiers, the noise performances is reported next. However notice that the matching and stability networks have been designed under the consideration of achieving maximum gain. The minimum noise figure $NF_{\rm min}$, noise resistance $R_{\rm n}$ and optimum reflection coefficient $\Gamma_{\rm opt}$ are of \SI{9.2}{\decibel}, \SI{103}{\Omega}, \SI{0.201}{} ($-\SI{101.1}{}^{\circ}$), respectively, for the PCA-PA and of \SI{6.7}{\decibel}, \SI{85}{\Omega}, \SI{0.213}{} ($\SI{4.3}{}^{\circ}$), respectively, for the PCA-IA. Low-noise amplifiers have been designed elsewhere \cite{RamPac19} with the same model and technology used here.

%\vspace{-0.6cm}
\subsection{Frequency configurable amplifier design}

The schematic of the FCA is similar to the PCA design, (Fig. \ref{fig:Pashe_Quadra}), however, parameter values are different (see Table I(b)).  Notice specially the inclusion of $R_{\rm D}$ in this design. The selected bias points are $V_{\rm GS}=\SI{0.3}{\volt}$ (region II) for the  frequency doubler operation mode and $V_{\rm GS}=\SI{0.9}{\volt}$ (region III) for the inverting amplifier operation mode, in both cases ${\textstyle V_{\rm DS}=\SI{3}{\volt}}$. The FCA name has been proposed since the output signal is at doubled or similar $f$ as the input signal, depending on the operation mode. The feedback network has been designed in order to have both stability conditions and an improved output signal in the FD operation mode. The matching network has been designed towards the enhancement of both operation modes. $P_{\rm in}$ is $-\SI{10}{\decibel m}$ at \SI{1.2}{\giga\hertz}.

\begin{figure}[!hbt]
	\centering
	{\includegraphics[height=0.22\textwidth]{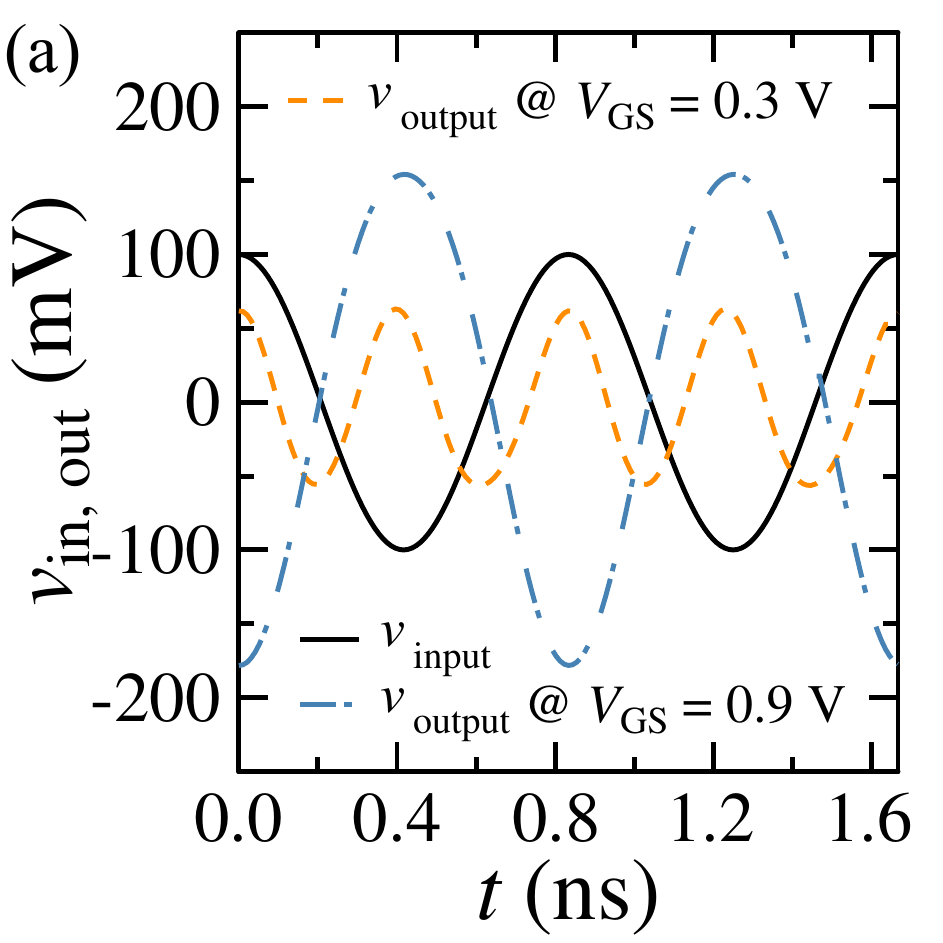}}
	{\includegraphics[height=0.22\textwidth]{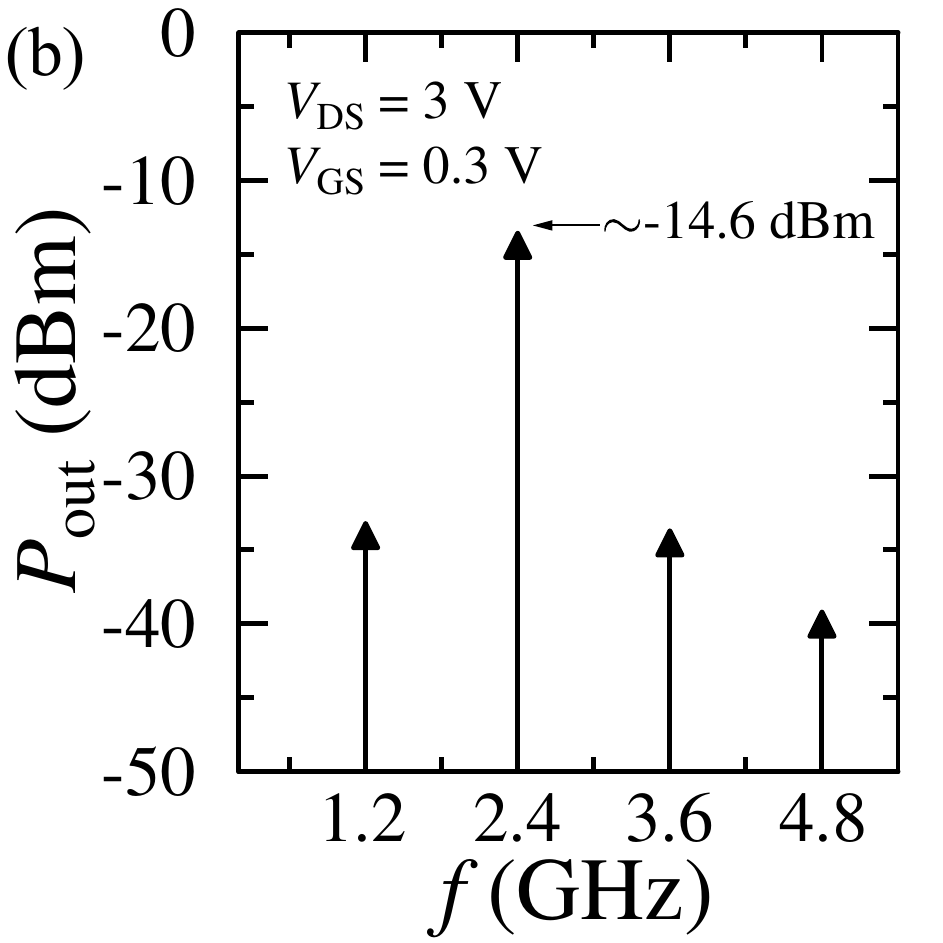}}
	\caption{(a) $v_{\rm in}$ and $v_{\rm out}$ signals for the two operation modes of the FCA: frequency doubler ($V_{\rm GS}=\SI{0.3}{\volt}$) and inverting amplifier ($V_{\rm GS}=\SI{0.9}{\volt}$) (a slight phase shift, caused by the matching networks, has been corrected in the plot for a better visualization of the circuit performance). (b) $P_{\rm out}$ spectrum of the frequency doubler operation mode for a $P_{\rm in}$ of $-\SI{10}{\decibel m}$ at \SI{1.2}{\giga\hertz}.}
	\label{fig:Voltaje_multiplicador}
\end{figure}

The frequency doubler performance ($V_{\rm GS}=\SI{0.3}{\volt}$), can be observed by comparing the input and output signals frequency shown in Fig. \ref{fig:Voltaje_multiplicador}(a). The slight difference in the amplitudes between two successive periods is related to a non-symmetric ambipolar device performance (see Fig. \ref{fig:Transfer}(a)). The reference device has not been optimized towards an ideal ambipolar symmetry since this technology has been proposed for general RF applications \cite{SchKol11}, e.g., \cite{EroLin11}-\cite{TagCar17}. The output power ($P_{\rm out}$) for the frequency doubler operation mode is $-\SI{14.6}{\decibel m}$ at \SI{2.4}{\giga\hertz} leading to a circuit conversion frequency loss of $\SI{4.6}{\decibel}$, while almost all the power is concentrated at \SI{2.4}{\giga\hertz} as shown in Fig. \ref{fig:Voltaje_multiplicador}(b). The FCA design is a more efficient frequency doubler than the reported circuits in \cite{WanLia14} with considerable lower conversion losses (-\SI{35}{\decibel} and -\SI{55}{\decibel}, regardless the different $n_{\rm t}$) at similar operation frequency. Additionally, the performance in $P_{\rm out}$ and operation frequency obtained here are higher than the ones of CNTFET-based frequency doublers presented elsewhere \cite{WanDin10}, \cite{LanYan20}. Furthermore, the voltage gain of the frequency doubler operation mode is of $\sim$\SI{0.6}{\volt/\volt} which, despite its low value, is almost \SI{4}{} times higher than the one achieved at \SI{2}{\kilo\hertz} with a different CNTFET technology \cite{WanZha13}. Multi-stage HF amplifiers solutions, with the same CNTFET technology \cite{ClaMuk14,RamPac19}, can be proposed towards improving the FCA-FD performance. 

\begin{figure}[!hbt]
	\centering
	{\includegraphics[height=0.22\textwidth]{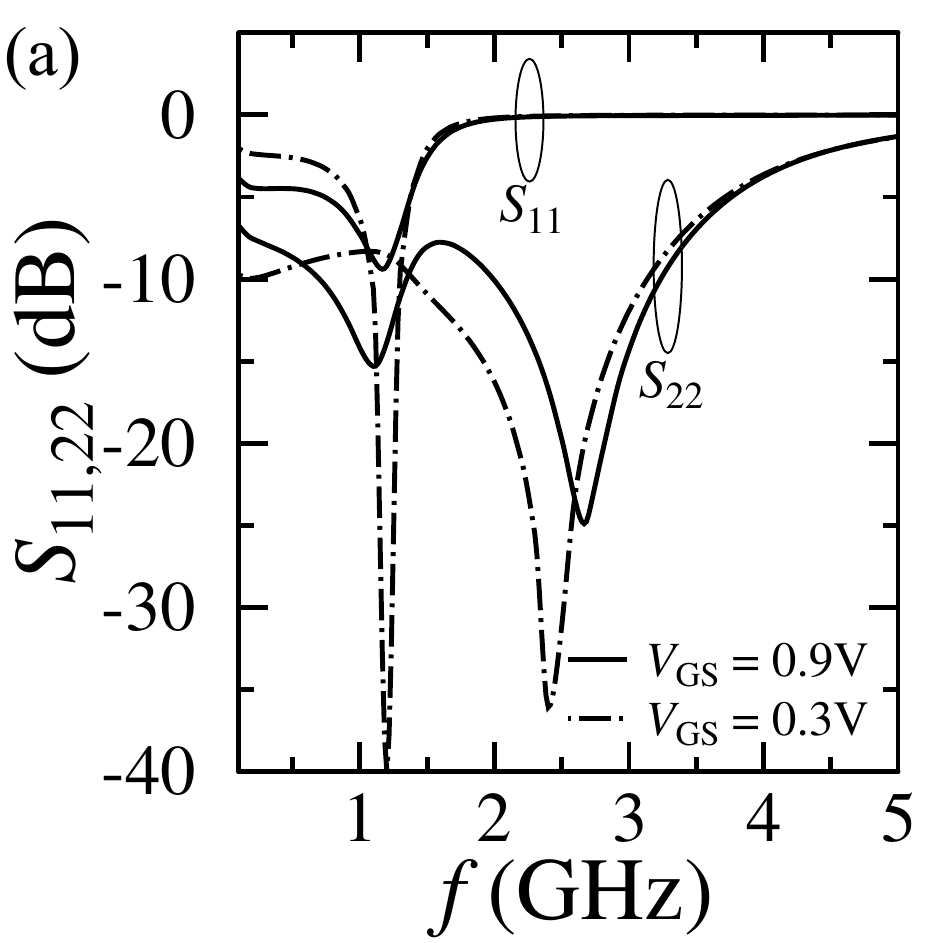}}
	{\includegraphics[height=0.22\textwidth]{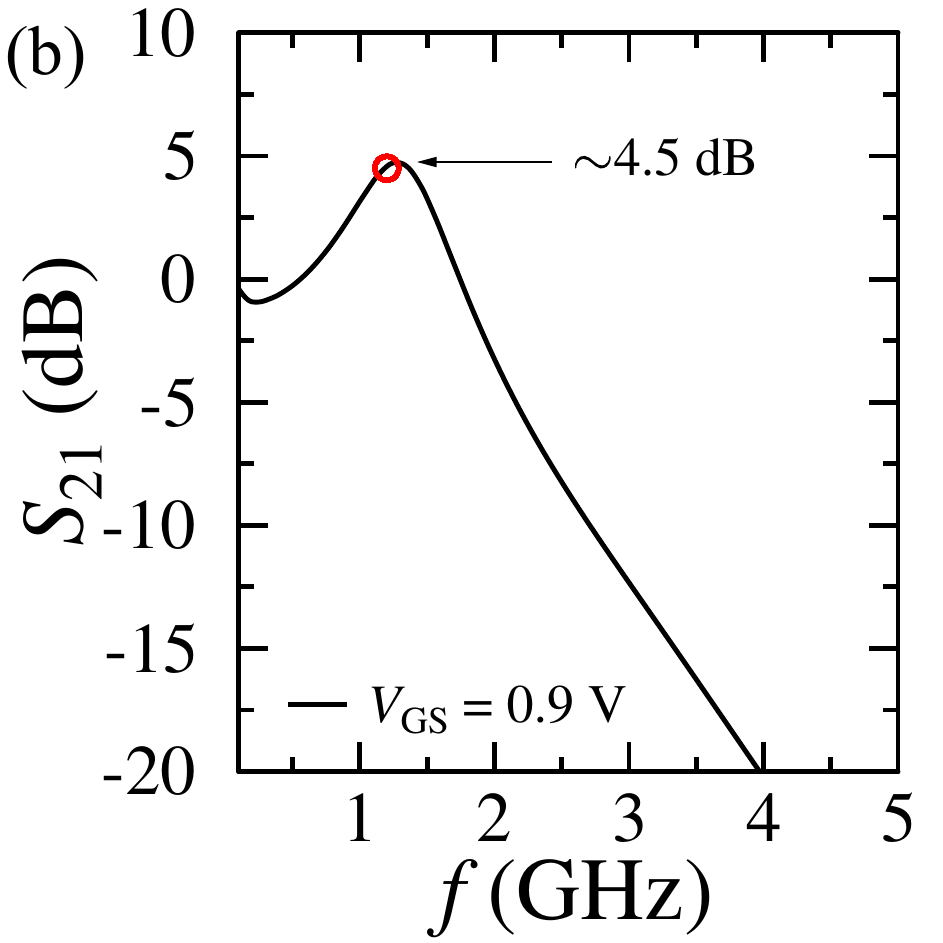}}
	\caption{(a) $\vert S_{11}\vert$ and $\vert S_{22}\vert$ for the two operation modes of the FCA. (b) $\vert S_{21}\vert$ for the FCA-IA (the circle indicates the gain at \SI{1.2}{\giga\hertz}).}
	\label{fig:K_multiplicador_ampli}
\end{figure}

Fundamental-harmonic suppression for the frequency doubler operation mode is \SI{19.65}{\decibel c} while the 3$^{\rm rd}$ and 4$^{\rm th}$ harmonic suppression are \SI{20.14}{\decibel c} and \SI{25.64}{\decibel c}, respectively. This indicates a high spectral purity for the doubled frequency $P_{\rm out}$. Notice that this harmonic suppression is higher than in previous CNTFET-related works \cite{WanLia14,WanDin10} where devices with high $n_{\rm t}$ have been considered. Device parasitics have been considered in the compact model, hence the achieved harmonic suppression is due to a proper matching network design, which produces $\vert S_{11}\vert$ and $\vert S_{22}\vert$ $<-\SI{10}{\decibel}$ at \SI{1.2}{\giga\hertz} and \SI{2.4}{\giga\hertz}, respectively (see Fig. \ref{fig:K_multiplicador_ampli}(a)), in addition to the device properties. Notice that the results obtained  for this circuit with this specific operation mode can be further improved if a more symmetric ambipolar response of the device can be achieved by, e.g., a modulation of the internal charge, however this is out of the scope of this study.

The FCA can be switched into amplification mode by biasing the device in region III ($V_{\rm GS}=\SI{0.9}{\volt}$). The device is stable at \SI{1.2}{\giga\hertz} and $\vert S_{11}\vert$ and $\vert S_{22}\vert$ $\lesssim$ $-\SI{10}{\decibel}$ at \SI{1.2}{\giga\hertz} as shows Fig.\ref{fig:K_multiplicador_ampli}(a). The unilateral power gain of the device in this bias point and at \SI{1.2}{\giga\hertz} is of \SI{7.4}{\decibel}. Power gain of $\sim$\SI{4.5}{\decibel} at \SI{1.2}{\giga\hertz} (see Fig.\ref{fig:K_multiplicador_ampli}(b)) is achieved with the FCA inverting amplifier operation mode. Noise figures of merit $NF_{\rm min}$, $R_{\rm n}$ and $\Gamma_{\rm opt}$) of the FCA-IA are \SI{10.3}{\decibel}, \SI{191.4}{\Omega} and \SI{0.186}{} ($-\SI{21.88}{}^{\circ}$), respectively. As discussed in Section \ref{ch:mul}.A, CNTFET-based amplifiers can be designed towards optimum noise performance as shown elsewhere \cite{RamPac19}. 

The performance of the FCA circuit can be controlled by the selected bias point without changes in the stability and matching networks resulting in a $V_{\rm GS}$-controlled circuit capable to operate as an amplifier at \SI{1.2}{\giga\hertz} or as a frequency doubler with an output $f$ of \SI{2.4}{\giga\hertz}. A third operation mode for the multifunctional circuits presented here, involves a more intricate trade-off between impedance matching and stability for each functionality and it is out of the scope of this work.

%\vspace{-0.3cm}
\section{Impact of m-tubes and inductor quality factor}

\begin{figure}[!hbt]
	\centering
	{\includegraphics[height=0.35\textwidth]{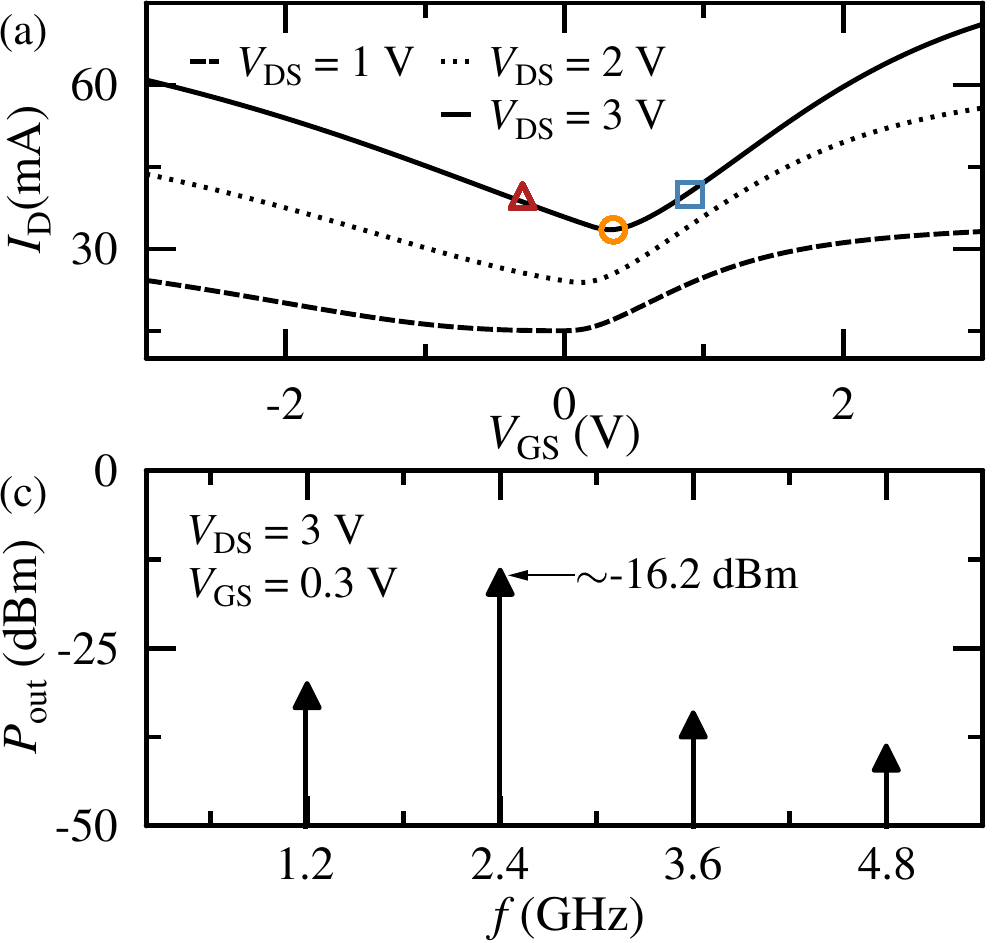}}
	{\includegraphics[height=0.35\textwidth]{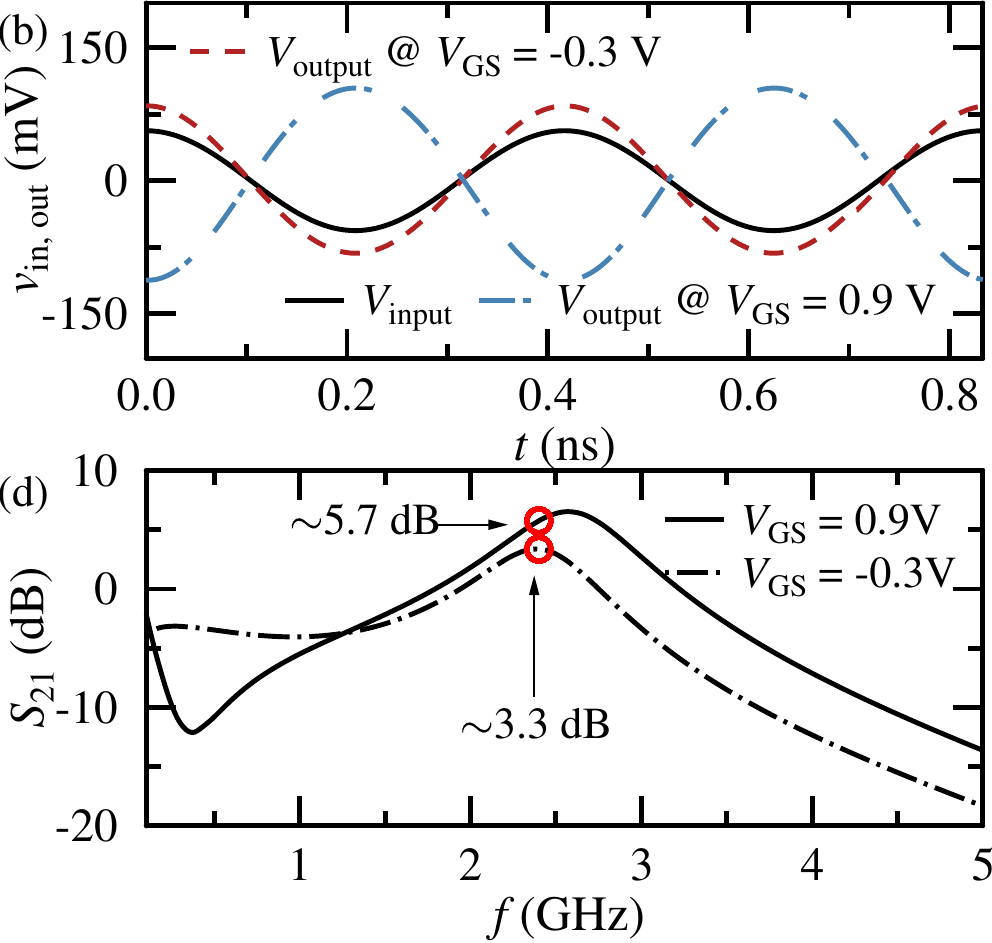}}
	\caption{Results with a s:m-ratio of 3:1 in the device channel: (a) CNTFET transfer characteristics, (b) $v_{\rm in}$ and $v_{\rm out}$ signals and (d) $\vert S_{21}\vert$ for the PCA operation modes, and (c) output spectrum of the FCA frequency doubler mode.}
	\label{fig:mCNTs}
\end{figure}

In order to consider a non-optimized techology, a study of the impact of m-tubes in the device channel on multifunctional circuits has been carried out here by considering a semiconducting-to-metallic tube ratio (s:m) of 3:1, i.e., the original calibration of the compact model (cf. Section \ref{ch:mod}) \cite{SchHaf15} has been used. Fig.  \ref{fig:mCNTs}(a) shows that device operation regions remain similar in shape with higher $I_{\rm D}$ values at the same bias in comparison to the optimal case (cf. Fig. \ref{fig:Transfer}(a)). At a circuit level however, higher $V_{\rm DD}$ are required for the FCA design (\SI{8.6}{\volt} for the FD and \SI{7.6}{\volt} for the IA). The overall performance of both multifunctional circuits is degraded with the presence of m-tubes (Figs. \ref{fig:mCNTs}(b)-(d)) in contrast to the CNTFET with s-tubes only, considered in previous sections. E.g., for the non-optimzed technology, gains of \SI{5.7}{\decibel} and \SI{3.3}{\decibel} have been obtained for the PCA-IA and PCA-PA, respectively, and a conversion loss of \SI{6.2}{\decibel} is achieved for the FCA-FD. The latter is obtained at the cost of more power consumption.

The impact of inductor quality factor ($Q$) on the performance of multifunctional circuits has been studied here by considering high-$Q$ on-chip CNT-based inductors \cite{LiBan09} in the proposed designs ($L_1$ in Fig. \ref{fig:Pashe_Quadra}). Results are shown in Table \ref{tab:Values_Q} for $Q$ of \SIlist{8;35}{} (see Fig. 12 in \cite{LiBan09}) as well as the ideal case of CNT inductors. All inductors can be assumed to be over SiO$_2$ substrates towards an integrated fabrication. The gain of the PCA's both modes is degraded $\sim\SI{3}{\decibel}$ in the worst scenario (lowest $Q$). Similarly, the FCA-IA gain is diminished $\sim\SI{1.5}{\decibel}$ for $Q=8$ while the FCA-FD conversion loss is $\sim\SI{3}{\decibel}$ more than the ideal case. In the case of $Q=35$ the performance is slightly deviated from the ideal case response.

%\vspace{-0.3cm}
\begin{table} [!hbt] 
	\centering 
	\caption{Impact of $Q$-factor on PCA and FCA operation modes at $\SI{2.4}{\giga\hertz}$}
	\subfloat[][PCA]{
		\centering
		\begin{tabular}{c|c||c}
			 & $Q$ & \makecell{$\vert S_{21}\vert$\\(\SI{}{\decibel})}  \\ \hline \hline
			
			\multirow{3}{*}{PA}      & 8     & \SI{1.4}{} \\
			                          & 35    & \SI{3.8}{}\\
                                      & ideal & \SI{4.5}{} \\ \hline
			\multirow{3}{*}{IA} & 8 & \SI{4}{} \\
			                          & 35 & \SI{6}{}  \\
                                      & ideal & \SI{6.7}{} \\                                      
		\end{tabular} 
	}
	\hspace{0.0mm}
	\subfloat[][FCA]{
		\centering
		\begin{tabular}{c|c||c|c}
			 & $Q$ & \makecell{$\vert S_{21}\vert$\\(\SI{}{\decibel})} & \makecell{conv. loss\\(\SI{}{\decibel})} \\ \hline \hline
			
			\multirow{3}{*}{FD} & 8 & -- & \SI{7.8}{} \\
			                       & 35 & -- & \SI{5.3}{} \\
                                   & ideal & -- & \SI{4.6}{} \\ \hline
			\multirow{3}{*}{IA} & 8     & \SI{3}{} & -- \\
			                    & 35    & \SI{4.2}{} & -- \\
                                & ideal & \SI{4.5}{} & -- \\            
            		\end{tabular}
	}
	\label{tab:Values_Q}
\end{table}
%\vspace{-0.6cm}

\section{CNTFET-based modulators enabled by the multifunctional circuits}

In order to show potential applications of the PCA and FCA with their different operation modes in high-data rate communication systems, two different modulation schemes have been demonstrated: phase-shift keying (PSK) and frequency-shift keying (FSK). The device compact model with semiconducting CNTs in the channel (Sections II and III) has been considered in this part of the work. 

\begin{figure}[!hbt]
	\centering
	{\includegraphics[height=0.225\textwidth]{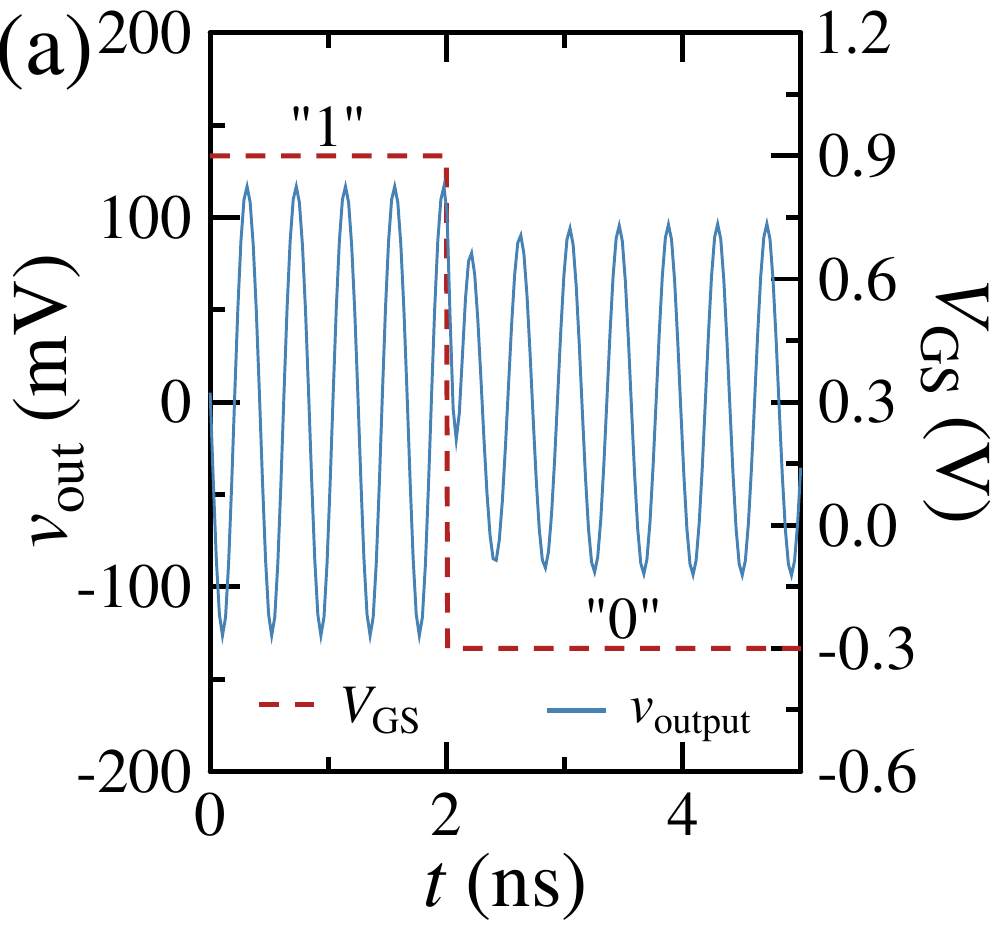}}
	{\includegraphics[height=0.225\textwidth]{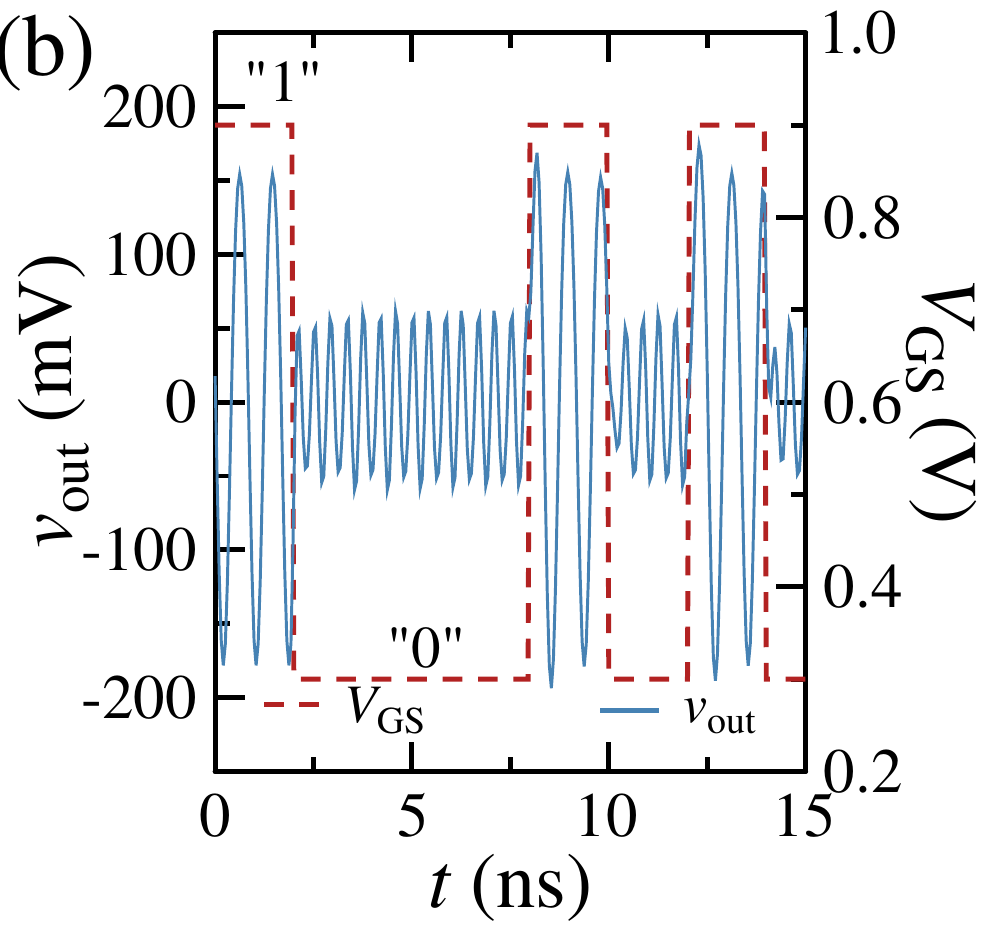}}
	\caption{Modulation schemes achieved with CNTFET-based multifucntional circuits. (a) PSK modlated signal ($v_{\rm out}$) achieved with the PCA by varying the operation mode between IA ($V_{\rm GS}=\SI{0.9}{\volt}$ equivalent to a digital "1") and PA ($V_{\rm GS}=-\SI{0.3}{\volt}$ equivalent to a digital "0"). (b) FSK modulated signal ($v_{\rm out}$) achieved with the FCA by varying the operation mode between IA ($V_{\rm GS}=\SI{0.9}{\volt}$ equivalent to a digital "1") and FD ($V_{\rm GS}=\SI{0.3}{\volt}$ equivalent to a digital "0").}
	\label{fig:modulators}
\end{figure}

The input has been fed with AC carrier signals (not shown here) with frequency of \SI{2.4}{\giga\hertz} for the PSK and of \SI{1.2}{\giga\hertz} for the FSK as well as the baseband data given by the variation of $V_{\rm GS}$. In all cases, $V_{\rm DS}=\SI{3}{\volt}$. By switching $V_{\rm GS}$ alternatively between \SI{0.9}{\volt} (IA operation mode) to \SI{-0.3}{\volt} (PA operation mode) the binary values "1" and "0" can be stablished with the PCA leading to a PSK modulated signal at the output as shown in Fig. \ref{fig:modulators}(a). Similarly, FSK modulation has been achieved by considering the two functionalities of the FCA as shown in Fig. \ref{fig:modulators}(b). Hence, the CNTFET-based circuit designs proposed in this work enable to perform modulation schemes, PSK with PCA or FSK with FCA, revealing them as suitable solutions for on-chip communication systems since only one single device and circuit have been used for each modulation. According to the authors' knowledge, this is the first proposal of CNTFET-based modulators.

\section{Conclusions}
\label{ch:con}

CNTFETs are promising candidates to develop multifunctional HF applications due to their intrinsic ambipolar transport. The circuits presented here include a PCA and a FCA. The PCA works at a $f$ of \SI{2.4}{\giga\hertz} for its two operation modes: as an in-phase amplifier and as an inverting amplifier. The FCA works at \SI{1.2}{\giga\hertz} for its two operation modes: as a frequency doubler and as an inverting amplifier. Similar matching networks have been used for both functionalities of each design. According to the authors' knowledge, this is the first proposal of CNTFET-based HF multifunctional circuits which configurability is only bias-controlled and without the need of changing matching and stabilization networks. These networks however, should be designed considering an enhanced performance for the two functionalities of each circuit. The results obtained here show the feasibility of multifunctional ambipolar CNTFET-based circuits for RF applications in S-band using a systematic study based on a device compact model calibrated with reproducible experimental data from a manufacturable CNTFET technology. This is of special interest towards multifunctional high-performance power-efficient on-chip solutions, e.g., in FSK or PSK modulators using a single device as shown in this work. Hence, this work's proposal can alleviate fabrication and production costs in HF integrated circuits based in CNTFETs or other ambipolar devices, e.g., graphene FETs.

In the case of the PCA, voltage and power gains ($\gtrsim$\SI{1.7}{\volt/\volt} and $\gtrsim$\SI{4.5}{\decibel}, respectively) have been achieved through a carefully bias point selection and stabilization and matching networks design. Moreover, fundamental-harmonic suppression close to \SI{20}{\decibel c} has been observed for the frequency doubler operation mode of the FCA, which indicates that doubler frequency circuits based on CNTFETs can be used without additional filtering stages. Moreover, both the impact of m-tubes in the device channel and CNTs-based inductors with different $Q$ on the circuits response have been studied. Results indicate that the multifuctionality feature of the designed circuits can be achieved with discrete success, in comparison to the optimized case, with at least $\SI{1}{}/\SI{3}{}$ of m-tubes in the channel array and with non-ideal CNT-based inductors.

%\vspace{-0.35cm}


\begin{thebibliography}{1}

\bibitem{Mar14} A. Marshall, ``Thoughts on Possible Future Charge-Based Technologies for Nano-Electronics", \emph{IEEE Trans. Circuits Syst. I, Reg. Papers}, vol. 61, no. 11, pp. 3057-3065, Jul. 2014. %DOI: 10.1109/TCSI.2014.2335011

\bibitem{HarHer21} M. Hartmann et al., "CNTFET Technology for RF Applications: Review and Future Perspective", \textit{IEEE Journal of Microwaves}, vol. 1, no. 1, pp. 275-287, Jan. 2021.

\bibitem{QiuZha17} C. Qiu, Z. Zhang, M. Xiao, Y. Yang, D. Zhong, L.-M. Peng, ``Scaling carbon nanotube complementary transistors to 5-nm gate lengths", \emph{Science}, vol. 355, pp. 271-276, Jan. 2017. %DOI: 10.1126/science.aaj1628

\bibitem{MotCla15} S. Mothes, M. Claus, and M. Schr\"oter, ``Towards linearity in Schottky-barrier CNTFETs", \emph{IEEE Trans. Nanotechnol.}, vol. 14, no. 2, pp. 372-378, Mar. 2015. %DOI: 10.1109/TNANO.2015.2397696.

\bibitem{Maa17} S. Maas, ``Linearity and Dynamic Range of Carbon Nanotube Field-Effect Transistors", in Proc. \emph{IEEE MTT-S International Microwave Symposium (IMS)}, Jun. 2017.% DOI: 10.1109/MWSYM.2017.8058727

\bibitem{CaoBra16} Y. Cao, G. J. Brady, H. Gui, C. Rutherglen, M. S. Arnold, C. Zhou, ``Radio Frequency Transistors Using Aligned Semiconducting Carbon Nanotubes with Current-Gain Cutoff Frequency and Maximum Oscillation Frequency Simultaneously Greater than 70 GHz", \emph{ACS Nano}, vol. 10, no. 7, pp. 6782-6790, Jun. 2016.% DOI: 10.1021/acsnano.6b02395

\bibitem{RutKan19} C. Rutherglen \emph{et al.}, "Wafer-scalable, aligned carbon nanotube transistors operating at frequencies of over 100 GHz", \emph{Nat. Electron.}, vol. 2, pp. 530-539, Nov. 2019.% DOI: 10.1038/s41928-019-0326-y

\bibitem{ZhoShi19} D. Zhong \emph{et al.}, "Carbon Nanotube Film-Based Radio Frequency Transistors with Maximum Oscillation Frequency above 100 GHz", \emph{ACS Appl. Mater. Interfaces}, vol. 11, no. 45, Oct. 2019.% DOI: 10.1021/acsami.9b15334

\bibitem{MarDer01} R. Martel \emph{et al.}, "Ambipolar Electrical Transport in Semiconducting Single-Wall Carbon Nanotubes", \emph{Phys. Rev. Lett.}, vol. 87, no. 25, 256805, Dec. 2001.% DOI: 10.1103/PhysRevLett.87.256805

\bibitem{RadHei03} M. Radosavljevic, S. Heinze, J. Tersoff, P. Avouris, ``Drain voltage scaling in carbon nanotube transistors", \emph{Appl. Phys. Lett.},  vol. 83, no. 12, 2435, Sep. 2003.% DOI: 10.1063/1.1610791

\bibitem{JavGuo04} A. Javey \emph{et al.}, "Carbon nanotube field-effect transistors with integrated ohmic contacts and high-$\kappa$ gate dielectrics", \emph{Nano Lett.}, vol. 4, no. 3, pp. 447-450, Feb. 2004.% DOI: 10.1021/nl035185x

\bibitem{ZhaWan09} Z. Zhang \emph{et al.}, ``Almost Perfectly Symmetric SWCNT-Based CMOS Devices and Scaling", \emph{ACS Nano}, vol. 3, no. 11, pp. 3781-3787, Oct. 2009.% DOI: 10.1021/nn901079p

\bibitem{BenJMoh11} M. H. Ben-Jamaa, K. Mohanram, G. De Micheli, ``An Efficient Gate Library for Ambipolar CNTFET Logic", \emph{IEEE Trans. Comput.-Aided Design Integr. Circuits Syst}, vol. 30, no. 2, pp. 242-255, Feb. 2011.% DOI: 10.1109/TCAD.2010.2085250

\bibitem{MouTie18} R. Moura, N. Tiencken, S. Mothes, M. Claus, S. Blawid, ``Reconfigurable NanoFETs: Performance Projections for Multiple-Top-Gate Architectures", \emph{IEEE Trans. Nanotechnol.}, vol. 17, no. 3, pp. 467-474, May 2018.% DOI: 10.1109/TNANO.2018.2812361

\bibitem{WanZha13} Z. Wang \emph{et al.}, ``Carbon Nanotube Based Multifunctional Ambipolar Transistors for AC Applications", \emph{Adv. Funct. Mater.}, vol. 23, pp. 446-450, 2013.% DOI: 10.1002/adfm.201202185

\bibitem{AppRad04} J. Appenzeller, M. Radosaveljevic, J. Knoch, P. Avouris, ``Tunneling Versus Thermionic Emission in One-Dimensional Semiconductors", \emph{Phys. Rev. Lett.}, vol. 92, 048301, Jan. 2004.% DOI: 10.1103/PhysRevLett.92.048301

\bibitem{PacFuc18} A. Pacheco-Sanchez \emph{et al.}, ``Feasible device architectures for ultra-scaled CNTFETs", \emph{IEEE Trans. Nanotechnol.}, vol. 17, no. 1, pp. 100-107, Jan. 2018.% DOI: 10.1109/TNANO.2017.2774605

\bibitem{LinApp04} Y.-M. Lin, J. Appenzeller, P. Avouris, ``Ambipolar-to-unipolar conversion of carbon nanotube transistors by gate structure engineering", \emph{Nano Lett.}, vol. 4, no. 5,  pp. 947-950, Mar. 2004.% DOI: 10.1021/nl049745j

\bibitem{JimCar06} D. Jim\'enez, X. Cartoix\`a, E. Miranda, J. Su\~n\'e, F. A. Chaves, S. Roche, ``A simple drain current model for Schottky-barrier carbon nanotube field effect transistors", \emph{Nanotechnology}, vol. 18, no. 2, Dec. 2006.% DOI: 10.1088/0957-4484/18/2/025201

\bibitem{CheLin13} Y. Che, Y.-C. Lin, P. Kim, C. Zhou, ``T-Gate Aligned Nanotube Radio Frequency Transistors and Circuits with Superior Performance", \emph{ACS Nano}, vol. 7 no. 5, pp. 4343-4350, Apr. 2013.% DOI: 10.1021/nn400847r

\bibitem{WanLia14} Z. Wang \emph{et al.}, ``Scalable Fabrication of Ambipolar Transistors and Radio-Frequency Circuits Using Aligned Carbon Nanotube Arrays", \emph{Adv. Mater.}, vol. 26, pp. 645-652, Jan. 2014.% DOI: 10.1002/adma.201302793

\bibitem{WanDin10} Z. Wang \emph{et al.}, ``Large Signal Operation of Small Band-Gap Carbon Nanotube-Based Ambipolar Transistor: A High-Performance Frequency Doubler", \emph{Nano Lett.}, vol. 10, pp. 3648-3655, Aug. 2010.% DOI: 10.1021/nl102111j

\bibitem{SchHaf15}M. Schr\"oter, M. Haferlach, A. Pacheco-Sanchez, S. Mothes, P. Sakalas, M. Claus, ``A Semiphysical Large-Signal Compact Carbon Nanotube FET Model for Analog RF Applications", \emph{IEEE Trans. Electron Devices}, Vol. 62, No. 1, pp. 52-60, Jan. 2015.% DOI: 10.1109/TED.2014.2373149


\bibitem{SchHaf15_2}M. Schr\"oter, M. Haferlach, M. Claus, ``CCAM Compact Carbon Nanotube Field-Effect Transistor Model (Version 2.0.3)", \emph{nanoHUB}.% DOI: 10.4231/D3VD6P595, 2015.

\bibitem{HafCla14} M. Haferlach, M. Claus, T. Nardmann, A. Pacheco-S\'{a}nchez, P. Sakalas, M. Schr\"{o}ter, ``Trap-induced Apparent Linearity of CNTFETs", \emph{Proc. Nanotech, Workshop on Compact Modeling (WCM)}, Washington D.C., U.S.A., Jun. 2014.

\bibitem{SchKol11}M. Schr\"oter \emph{et al.}, ``A 4" Wafer Photostepper-Based Carbon Nanotube FET Technology for RF Applications", \emph{Proc. IEEE MTT-S International Microwave Symposium}, Jun. 2011.% DOI: 10.1109/MWSYM.2011.5972750

\bibitem{ClaMuk14} M. Claus, A. Mukherjee, A. Moroguma, A. Pacheco, S. Blawid, M. Schr\"oter, ``High frequency benchmark circuit design for a sub-50nm CNTFET technology", \emph{Proc. SBMO-IEEE MTT-S International Microwave and Optoelectronics Conference (IMOC)}, Rio de Janeiro, Brazil, Aug. 2013.% DOI: 10.1109/imoc.2013.6646521

\bibitem{RamPac19} J. N. Ramos-Silva, A. Pacheco-Sanchez, L. M. Diaz-Albarran, L. M. Rodriguez-Mendez, M. Schr\"oter, E.  Ramirez-Garcia, ``High-Frequency Performance Study of CNTFET-Based Amplifiers", \emph{IEEE Trans. Nanotechnol.}, vol. 19, pp. 284-291, Jan. 2020.% doi: 10.1109/TNANO.2020.2978816.

\bibitem{EroLin11}M. Eron, S. Lin, D. Wang, M. Schr\"oter, P. Kempf, ``L-band carbon nanotube transistor amplifier", \emph{Electron. Lett.}, vol. 47, no. 4, pp. 265-266, Feb. 2011.% DOI: 10.1049/el.2011.0018

\bibitem{TagCar15}A. Taghavi, C. Carta, F. Ellinger, M. Haferlach, M. Claus, M. Schr\"oter, ``A CNTFET Amplifier with 5.6 dB Gain Operating at 460-590 MHz", \emph{Proc. SBMO/IEEE MTT-S International Microwave and Optoelectronics Conference (IMOC)}, Nov. 2015.% DOI: 10.1109/IMOC.2015.7369110

\bibitem{TagCar17}A. Taghavi, C. Carta, T. Meister, F. Ellinger, M. Claus, M. Schr\"oter, ``A CNTFET Oscillator at 461 MHz", \emph{IEEE Microw. Wireless Compon. Lett.}, vol. 27, no. 6, pp. 578-580, May. 2017.% DOI: 10.1109/LMWC.2017.2701312

\bibitem{CaoChe16}Y. Cao, Y. Che, H. Gui, X. Cao, C. Zhou, ``Radio Frequency Transistors Based on Ultra-High Purity Semiconducting Carbon Nanotubes with Superior Extrinsic Maximum Oscillation Frequency", \emph{Nano Res.}, vol. 9, no. 2, pp. 363-371, Nov. 
2016.% DOI: 10.1007/s12274-015-0915-7

\bibitem{Poz11}D. M. Pozar, ``Microwave Engineering", John Wiley and Sons, 2011 (USA), Chap. 12.

\bibitem{LanYan20} Y. Lan et al., "High-Temperature-Annealed Flexible Carbon Nanotube Network Transistors for High-Frequency Wearable Wireless Electronics", \textit{ACS Applied Materials \& Interfaces}, vol. 12, pp. 26145-26152, May 2020.

\bibitem{LiBan09} H. Li and K. Banerjee, ``High-frequency analysis of carbon nanotube interconnects and implications for on-chip inductor design", \emph{IEEE Trans. Electron Devices}, vol. 56, no. 10, pp. 2202–2214, Oct. 2009.


\end{thebibliography}
\end{document}